\documentclass[a4paper,11pt]{article}
\usepackage[usenames]{color}
\usepackage[dvips]{graphicx}
\usepackage{graphicx}
\usepackage{epsf}
\usepackage{amsmath}

\newcommand{\Pmatrix}[1]{\begin{pmatrix} #1 \end{pmatrix}}

\newcommand{\bea}{\begin{eqnarray}}
\newcommand{\eea}{\end{eqnarray}}

\newcommand{\IkB}{I$\kappa$B }
\newcommand{\NFkB}{NF-$\kappa$B }

\begin{document}

\begin{titlepage}
\title{Stochastic Binary Modeling of Cells in Continuous Time\\ as an Alternative to Biochemical Reaction Equations}
\author{\vspace{2mm}
Shunsuke Teraguchi$^1$, Yutaro Kumagai$^1$, Alexis Vandenbon$^2$,\\ Shizuo Akira$^1$ and Daron M Standley$^2$\\[10mm]
$^1$\textit{\normalsize Laboratory of Host Defense,}
$^2$\textit{\normalsize Laboratory of Systems Immunology,}\\
\textit{\normalsize WPI Immunology Frontier Research Center (IFReC), Osaka University,}\\
\textit{\normalsize 3-1 Yamada-oka, Suita, Osaka 565-0871, Japan}
}
\date{}
\maketitle
\thispagestyle{empty}
\begin{abstract}
We have developed a coarse-grained formulation for modeling the dynamic behavior of cells quantitatively,
based on stochasticity and heterogeneity, rather than on biochemical reactions.
We treat each reaction as a continuous-time stochastic process,
while reducing each biochemical quantity to a binary value at the level of individual cells.
The system can be analytically represented by a finite set of ordinary linear differential equations,
which provides a continuous time course prediction of each molecular state.
In this letter, we introduce our formalism and demonstrate it with several examples.
\end{abstract}
\end{titlepage}

\section{Introduction}
With the rapid growth in molecular biology and its related fields,
there is a need for a quantitative theoretical framework \cite{de2002modeling} that can both integrate biological knowledge and provide useful predictions to guide further experiments. 
The most standard approach for describing biological phenomena at the molecular level is through the use of reaction equations, which originate from chemistry.
If the rates, stoichiometries, and initial conditions of each species are known,
one can in principle model arbitrarily complex processes inside a cell.
Even without such detailed information, this approach is still very advantageous if the system is approximately static, most of the stoichiometries and conserved quantities are known, and a suitable objective function can be defined, as in metabolic networks \cite{edwards2001silico}.
On the other hand, for systems without these conditions, such as signal transduction networks, precise modeling of biological reactions remains challenging.
In practice, there are three major obstacles to this approach.
The first results from incompleteness of our knowledge at a molecular level: Accurate modeling requires correct stoichiometries and reaction constants, not to mention knowledge of all the reactions involved in a particular reaction network.
With very few exceptions \cite{hoffmann2002ikb}, we lack such knowledge.
The second is the inherent heterogeneity of living cells: Molecular biology experiments are generally performed using millions of cells at a single time point; it is not clear whether measured values reflect the state of a typical cell or an average of distributions of cells whose states are largely different from each other \cite{raj2008nature,tay2010single}.
The third is the difficulty in obtaining absolute concentrations from such experiments: Most standard biochemical experiments measure only relative concentrations of molecules with respect to a standard; converting these relative values into absolute concentrations requires further experiments.

The Boolean approach \cite{kauffman1969metabolic}, in which all the states of molecules are represented by True of False binary values and Boolean algebra is used to define relationships between molecules, is another popular approach to circumvent some of the above obstacles. Despite its simplicity, the Boolean approach has been successfully applied to biological systems \cite{morris2010logic}. However, the price we pay for these simplifications is that models largely remain qualitative.

On the other hand, recent studies have revealed that the stochastic and heterogeneous nature of cells is indispensable to understanding the design of the cellular dynamics \cite{kaern2005stochasticity,bala'zsi2011cellular}. 
In this paper, we propose a stochastic generalization of the Boolean approach as an alternative to biochemical reaction equations.
In the following, we shall see how this formulation enables us to describe the dynamics of cellular systems
while circumventing the above problems on biochemical reaction equations.
Although this formulation would be applicable to various biological phenomena, we explain it in application to signal transduction or gene regulation.

\begin{figure} [htbp]
\centering
\includegraphics[width=0.5\linewidth]{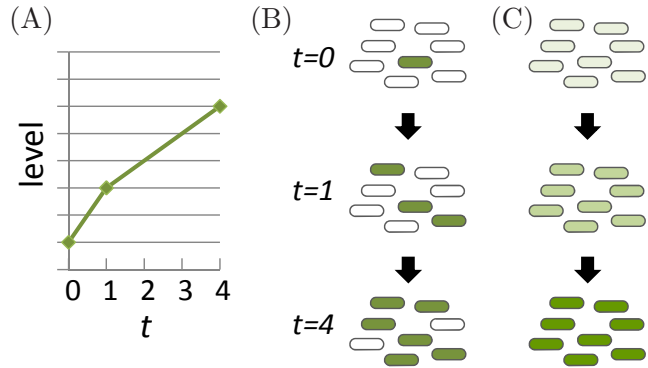}
\put(-240,130){(A)}
\put(-150,130){(B)}
\put(-60,130){(C)}
\caption{A schematic representation of two possible approximations for interpreting bulk assays. \label{illust}
(A) A typical output of a bulk time course experiment.
(B) Cells behave stochastically and digitally, as assumed in our formulation. 
(C) Every cell behaves coherently and continuously, as assumed in the deterministic biochemical reaction modeling.}
\end{figure}

\section{Formulation and Examples}
It is convenient to represent such a system by a regulatory network diagram of biochemical species (nodes) and interactions (edges) between them.
We use the term "level" to represent the value associated with each node, which may represent concentrations of molecules, expression levels of genes, enzyme activities and so on. 
While the deterministic biochemical reaction models implicitly assume that the observed levels reflect the concentration of the molecular species in a typical cell, we assume that experimental observations on large numbers of cells reflect averages of distributions of cells whose states can vary significantly (FIG.\ \ref{illust}).
Though, in reality, each cell would take various values for each node, we approximate them by a binary value, True ($T$) or False ($F$), as in the Boolean approach. 
Thus, though there are several similarities to earlier attempts to introduce randomness into Boolean models \cite{shmulevich2002probabilistic},
our approach is intrinsically related to the heterogeneous nature of cells.
We define interactions between the states of individual molecules by stochastic processes in continuous time. Here, in order to facilitate mathematical analysis, we make the assumption that they satisfy the Markov property.

\subsection{Simple systems with two nodes}
\begin{figure} [htbp]
\centering
\begin{minipage}{0.12\linewidth}
\includegraphics[width=\linewidth]{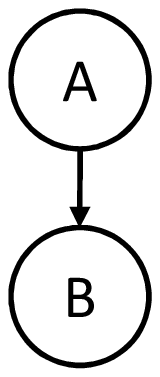}
\put(-50,80){(A)}
\put(-38,35){$\tau$}
\end{minipage}
\phantom{aa}
\leavevmode
\put(-7,-20){\rotatebox{90}{\footnotesize The level of $B$}}
\put(75,-52){\footnotesize time}
\begin{minipage}{0.32\linewidth}
\includegraphics[width=\linewidth]{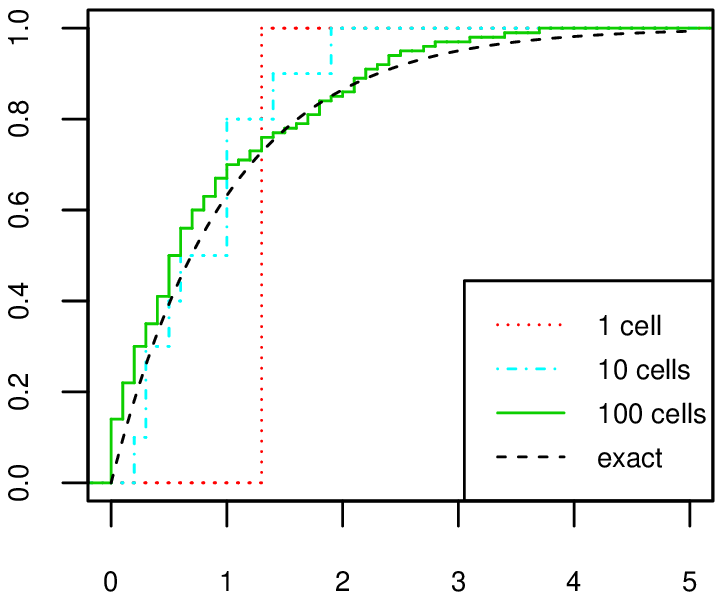}
\end{minipage}
\phantom{aaa}
\leavevmode
\begin{minipage}{0.12\linewidth}
\includegraphics[width=\linewidth]{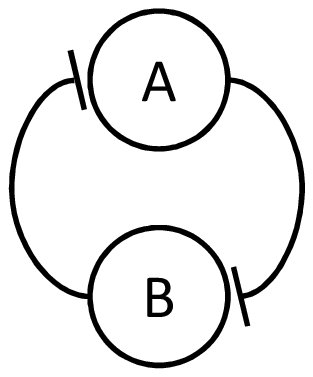}
\leavevmode
\put(-50,80){(B)}
\put(-50,35){$\tau_2$}
\put(-15,35){$\tau_1$}
\end{minipage}
\phantom{aa}
\leavevmode
\put(-7,-20){\rotatebox{90}{\footnotesize The level of $B$}}
\put(75,-52){\footnotesize time}
\begin{minipage}{0.32\linewidth}
\includegraphics[width=\linewidth]{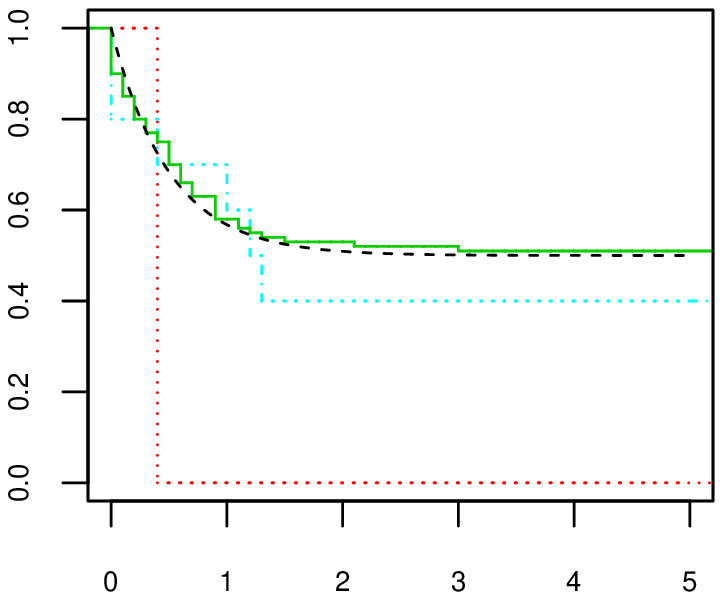}
\end{minipage}
\caption{Examples of systems with two nodes. \label{simple_reg}
(A) A system with an interaction. The initial condition is $P_{TF}=1$.
The black dashed line in the plot shows the exact solution of the master equation ($\tau=1$), while the others are the result of Monte-Carlo simulations performed 1, 10 and 100 times, respectively.
(B) A bistable system and the corresponding plots ($\tau_1=\tau_2=1$), as in (A).}
\end{figure}

To illustrate how our method works,
let us first consider the simplest case where an interaction is defined between two single nodes (FIG.\ \ref{simple_reg}A).
The time-scale parameter $\tau$ on the edge represents the typical time-scale of the interaction.
Here, the arrow from node A to node B means ``if A is active, B will be activated at the stochastic rate of $1/\tau$''. We may represent the same information by the following equation:
\begin{equation}
A\stackrel{\tau}{\longrightarrow} B.
\end{equation}
One can perform Monte Carlo simulation using methods such as the Gillespie algorithm \cite{gillespie1977exact} to realize the dynamics as in FIG.\ \ref{simple_reg}.
Here, the term ``cells'' represents the number of independent simulations.
While each cell behaves digitally, averaging over multiple cells gives more smooth and deterministic behavior.
Indeed, in the limit of the cell number approaching infinity,
this system can be analytically represented by the so-called master (or Kolmogorov) equation:
\begin{equation}
\frac{d}{dt}
\Pmatrix{
P_{TF}\\
P_{TT}\\
P_{FT}\\
P_{FF}
}(t)=
\Pmatrix{
-1/\tau &0&0&0\\
1/\tau &0&0&0\\
0&0&0&0\\
0&0&0&0\\
}
\Pmatrix{
P_{TF}\\
P_{TT}\\
P_{FT}\\
P_{FF}
}(t),
\end{equation}
where $P_{ij}(t)$ for $i,j\in\{T,F\}$ represents the probability of a cell taking the state of $(A=i,B=j)$ at time $t$, and the matrix appearing in the right hand side is the transition rate matrix of the system.
The dynamics of the system is fully encoded in the solution:
\bea
P_{TF}(t)&=&\exp\left(-\frac{t}{\tau}\right)P_{TF}(0),\\
P_{TT}(t)&=&\left(1-\exp\left(-\frac{t}{\tau}\right)\right)P_{TF}(0)+P_{TT}(0),\\
P_{FT}(t)&=&P_{FT}(0),\\
P_{FF}(t)&=&P_{FF}(0).
\eea
Now, imagine that we are measuring the level of $B$ by a bulk assay.
Then the observed level should be linearly related to the number of cells with $B=T$, 
in other words, the  marginal probabilities of $B=T$,
\begin{equation}
P_{*T}(t)\equiv P_{FT}(t)+P_{TT}(t)=\left(1-\exp\left(-\frac{t}{\tau}\right)\right)P_{TF}(0)+P_{*T}(0).
\end{equation}
Let us assume $l_T$ ($l_F$) is the observed level when all the cells at the measurement are in the activated (inactivated) state, respectively.
Then, our prediction for the level of $B$ would be given by
\begin{equation}
l_B(t)\equiv l_T P_{*T}(t)+l_F(1-P_{*T}(t))=(l_T-l_F) P_{*T}(t)+l_F.
\end{equation}
Thus, 
the prediction for each level is given by a linear transformation of the corresponding marginal probability.
Note that, only the two parameters $l_T$ and $l_F$ for each node are responsible for the scale of the level of the node.
This is in contrast to biochemical reaction equations where a change in the scale of each node non-linearly affects the scales and dynamics of all the other nodes.
For observations of relative values with respect to a standard,
these scaling parameters can be conveniently used for absorbing the missing information.
In the rest of this paper, the translation from probability to observed data is implicit and not mentioned.
Note also that in our formulation, although we approximate the possible states of each node by a binary value as in Boolean models, the model produces continuous time course data on the levels of the molecules because
the final prediction is obtained by averaging over the population.\\

The second example (FIG.\ \ref{simple_reg}B) is a mutual inhibition:
\bea
A\stackrel{\tau_1}{\longrightarrow}\ !B,\\
B\stackrel{\tau_2}{\longrightarrow}\ !A,
\eea
which is expected to show bistability \cite{gardner2000construction}.
Here the symbol ``$!$'' in front of the nodes in the right hand side indicates that the node will be inactivated, rather than activated, at the indicated rate.
The master equation for the system is given by
\begin{equation}
\frac{d}{dt}
\Pmatrix{
P_{TF}\\
P_{TT}\\
P_{FT}\\
P_{FF}
}(t)=
\Pmatrix{
0&1/\tau_1&0&0\\
0&-1/\tau_1-1/\tau_2&0&0\\
0&1/\tau_2&0&0\\
0&0&0&0\\
}
\Pmatrix{
P_{TF}\\
P_{TT}\\
P_{FT}\\
P_{FF}
}(t),
\end{equation}
and the solution with the initial condition of $P_{TT}=1$ is
\bea
P_{TF}(t)&=&\frac{\tau_2}{\tau_1+\tau_2}\left(1-\exp\left(-\left(\frac{1}{\tau_1}+\frac{1}{\tau_2}\right)t\right)\right),\\
P_{TT}(t)&=&\exp\left(-\left(\frac{1}{\tau_1}+\frac{1}{\tau_2}\right)t\right),\\
P_{FT}(t)&=&\frac{\tau_1}{\tau_1+\tau_2}\left(1-\exp\left(-\left(\frac{1}{\tau_1}+\frac{1}{\tau_2}\right)t\right)\right).
\eea
In sharp contrast to the solution of the deterministic biochemical reaction equations,
where only one of two stable states is taken depending on the initial condition,
our formulation naturally describes the situation where
a portion of the cell population flows to one state and the rest goes to the other.
This is because our formulation takes the heterogeneity of cell culture into account from the beginning.

\subsection{A simple oscillatory system}
It is also interesting to consider how this formulation describes oscillatory behavior,
which is prevalent in many biological systems, including cell cycle, circadian rhythm or calcium signaling.
A prototype for such oscillatory systems in Boolean models might be the following negative feedback system with two nodes \cite{ferrell2011modeling}:
\bea
A&{\longrightarrow}&B,\\
B&{\longrightarrow}&!A,\\
!A&{\longrightarrow}&!B,\\
!B&{\longrightarrow}&A.
\eea
Here the symbol ``$!$'' in front of the nodes in the left hand side indicates that the interaction occurs when the corresponding node is not activated.
This system itself oscillates.
However, it is not obvious whether the oscillation is observed at the population level or not.
Assuming a common time-scale parameter $\tau$ for each interaction, the master equation for this system is given by
\begin{equation}
\frac{d}{dt}
\Pmatrix{
P_{TF}\\
P_{TT}\\
P_{FT}\\
P_{FF}
}(t)=
\Pmatrix{
-1/\tau&0&0&1/\tau\\
1/\tau&-1/\tau&0&0\\
0&1/\tau&-1/\tau&0\\
0&0&1/\tau&-1/\tau\\
}
\Pmatrix{
P_{TF}\\
P_{TT}\\
P_{FT}\\
P_{FF}
}(t),
\end{equation}
and the marginal probabilities are found to be
\begin{equation}
P_{T*}(t)=\left(1+e^{-\frac{t}{\tau}}\left(\sin\left(\frac{t}{\tau}\right)+\cos\left(\frac{t}{\tau}\right)\right)\right)/2
\end{equation}
and
\begin{equation}
P_{*T}(t)=\left(1+e^{-\frac{t}{\tau}}\left(\sin\left(\frac{t}{\tau}\right)-\cos\left(\frac{t}{\tau}\right)\right)\right)/2
\end{equation}
under the initial condition of $P_{TF}=1$.
While the presence of the trigonometric functions indicates some oscillatory behavior,
they are multiplied by a rapidly decaying factor $e^{-\frac{t}{\tau}}$.
Therefore, on the population level, the oscillation is quickly extinguished due to decoherence between cells. 
In particular, under the assumption of the Markov property, where the timing of the processes follows an exponential distribution,
the standard deviation is as large as the mean time\footnote{Note also that one can achieve smaller standard deviations by connecting multiple nodes in series while keeping the same total time-scales.}
$\tau$
and the decoherence occurs very quickly.
Thus, even though each cell behaves in complicated ways,
these independent behaviors are averaged at the population level and not observed in the final prediction.
This consideration also provides an intuitive interpretation of the linearity of our formalism.
While it is widely believed that biological systems are full of non-linearities,
the master equation, whose solutions are related to observed values in this formulation,
is always linear and robust, and shows neither chaotic nor divergent behavior in any parameter region.

\subsection{Boolean algebra}
For more complicated systems beyond two nodes,
the formulation is quite parallel, 
though the number of possible states becomes large for systems with many nodes.
The only new ingredients are the variety of possible interactions.
With multiple nodes, one can consider interactions which depend on a set of input nodes.
For example, one may introduce the ``and'' operation: 
\begin{equation}
A\&B\stackrel{\tau}{\longrightarrow}C.
\end{equation}
Here, this equation means ``if both $A$ and $B$ are active, $C$ will be activated at the stochastic rate of $1/\tau$''
and one can easily write down the corresponding master equation.
In a similar way, arbitrary Boolean operation can be adopted.

\begin{figure*} [htbp]
\centering
\begin{minipage}{0.22\linewidth}
\includegraphics[width=\linewidth]{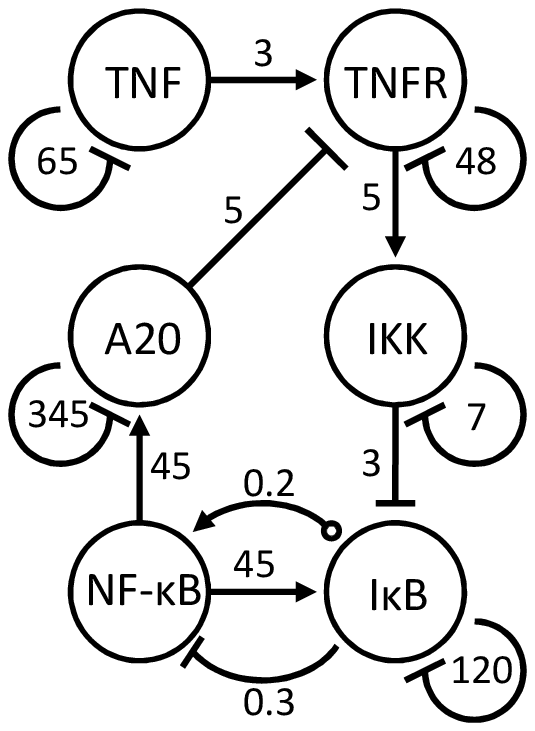}
\leavevmode
\put(-110,160){(A)}
\end{minipage}
\phantom{a}
\leavevmode
\put(-10,80){(B)}
\put(39,-84){\footnotesize time (min)}
\put(134,-84){\footnotesize time (min)}
\begin{minipage}{0.40\linewidth}
\includegraphics[width=\linewidth]{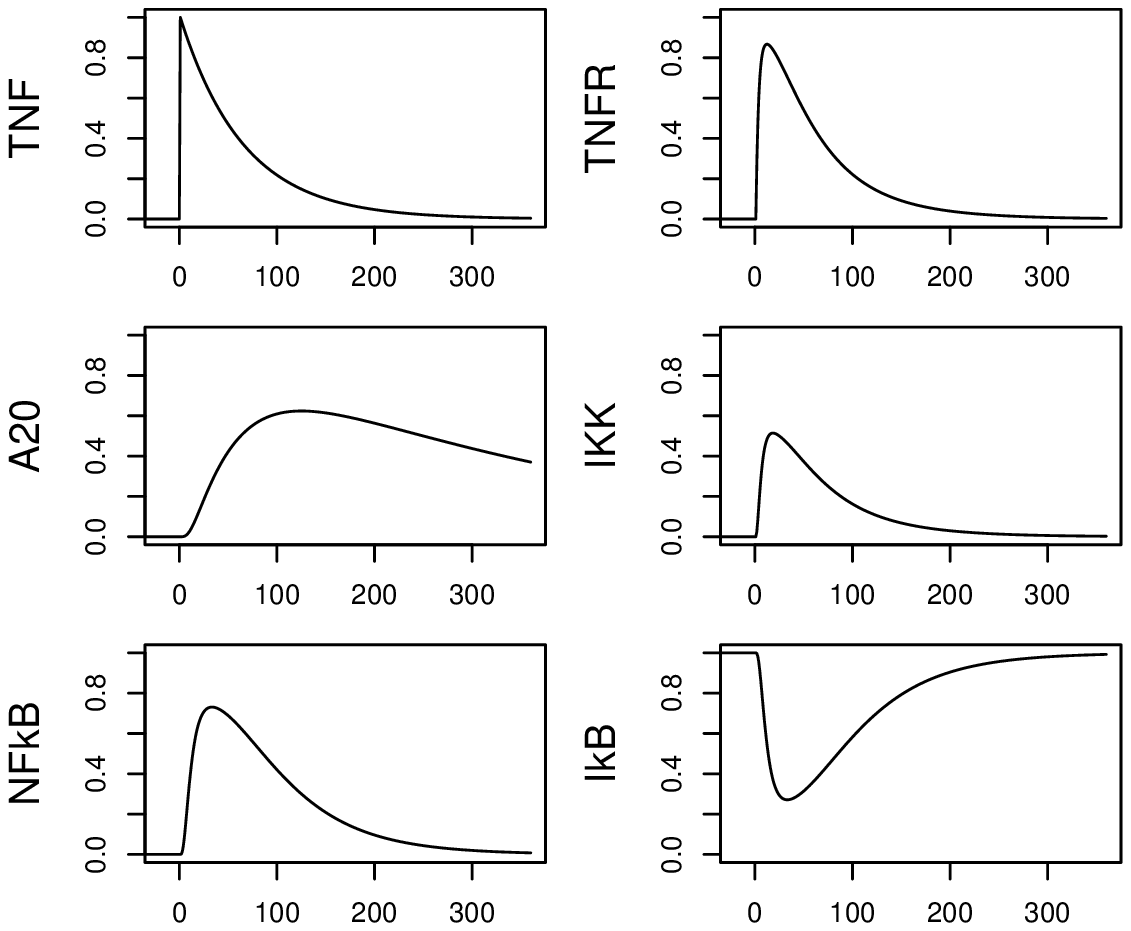}
\end{minipage}
\leavevmode
\put(5,80){(C)}
\phantom{aa}
\begin{minipage}{0.27\linewidth}
\leavevmode
\put(-7,23){\rotatebox{90}{\footnotesize NF-$\kappa$B activation}}
\put(47,0){\footnotesize time (min)}
\includegraphics[width=\linewidth]{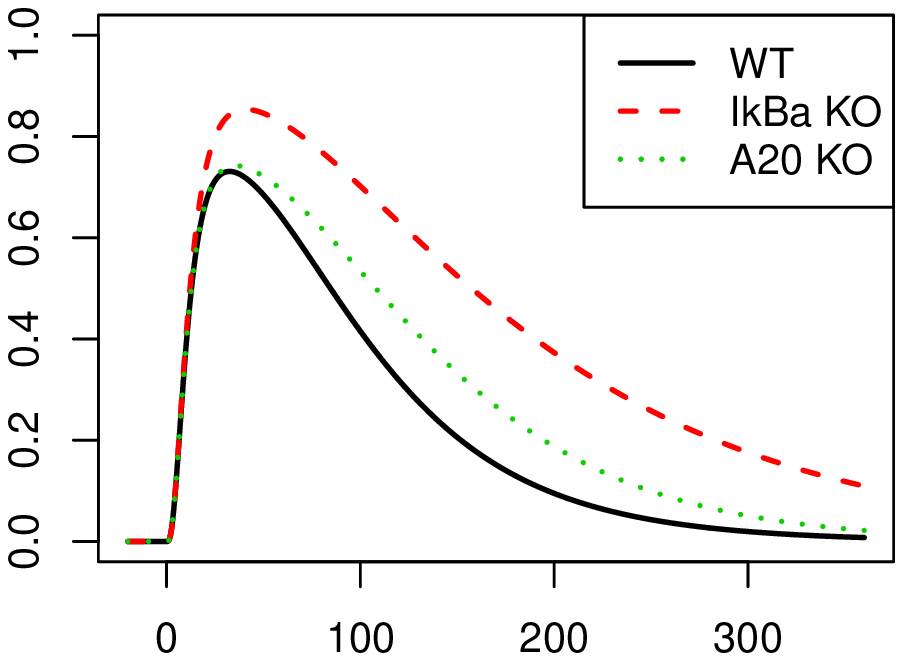}
\end{minipage}
\vspace{-3mm}
\caption{A model of the TNF--NF-$\kappa$B system.
(A) The network diagram. The number on each edge is the time-scale parameter ($\min$) for the interaction.
The small circle at the end of the arrow from I$\kappa$B to NF-$\kappa$B indicates that the interaction occurs when the I$\kappa$B is inactivated.
(B) The predicted time courses of our model for WT.
(C) The change of the NF-$\kappa$B activation level for WT, I$\kappa$B KO and A20 KO.
\label{NF-kB}}
\end{figure*}

\subsection{A demonstration: TNF--\NFkB system}
As discussed above,
the aim of our formulation is to model dynamic cellular systems in a coarse-grained way.
As a proof of concept, we next consider a signaling pathway which has previously been studied by biochemical reaction equations.
Werner et al.\ modeled the TNF--NF-$\kappa$B system  using biochemical reaction equations \cite{hoffmann2002ikb,werner2008encoding}.
TNF--NF-$\kappa$B pathway is an important signaling pathway in immune cells,
which can be activated by the cytokine TNF,
and negatively regulated by I$\kappa$B and A20.
The biochemical reaction equation model consisted of 33 molecular species and 110 kinetic coefficients.
In contrast, as shown in FIG.\ \ref{NF-kB}A, our model consists of only 6 nodes and 13 time-scale parameters\footnote{See Appendix for details of the system and parameters.}.
Before stimulation, the system is in a steady state, with the transcription factor NF-$\kappa$B inactivated by the inhibitor protein I$\kappa$B.
Once the system is stimulated by the ligand TNF, the TNFR signaling complex is formed.
This complex formation leads to activation of the kinase IKK. IKK then phosphorylates I$\kappa$B, leading to its degradation.
Since I$\kappa$B is an inhibitor of NF-$\kappa$B, I$\kappa$B degradation results in NF-$\kappa$B activation.
NF-$\kappa$B then upregulates the transcription of various genes, including I$\kappa$B and A20. I$\kappa$B, in turn, inhibits NF-$\kappa$B and A20 inhibits the TNFR complex.
As shown in FIG.\ \ref{NF-kB}B, our model predicted the time-dependent levels of each species in the network.
The authors of \cite{werner2008encoding} examined their model assuming wild-type (WT) conditions,
I$\kappa$B deficient conditions, (I$\kappa$B KO), and A20 deficient conditions (A20 KO): In the absence of I$\kappa$B, NF-$\kappa$B activation overshoots, whereas A20 deficiency leads to delayed adaptation of NF-$\kappa$B \cite{werner2008encoding}. 
In order to validate our model, we also suppressed I$\kappa$B induction (deleting ``NF-$\kappa$B $\stackrel{45}{\longrightarrow}$ I$\kappa$B''; I$\kappa$B KO) and A20 induction (deleting ``NF-$\kappa$B $\stackrel{45}{\longrightarrow}$ A20''; A20 KO).
As shown in FIG.\ \ref{NF-kB}C, the resulting time-course levels of NF-$\kappa$B reproduced the experimentally observed phenotypes.
Thus, although our formulation does not depend on the details of the biochemical reactions,
it still provides an alternative method for modeling signaling pathways.

\section{Summary and Discussion}
In this paper, we have presented a novel formulation of cellular processes at the population level,
based on a discrete and stochastic description at the single cell level.
Though most of the examples here are simple toy models,
our motivation is toward application to more realistic and complicated systems of cells such as signal transduction and transcriptional regulatory networks, as demonstrated in the final example.
To this end we constructed a model of the TNF--\NFkB system using only the coarse-grained topology and typical time-scale parameters of the system.
The fact that our models were built without fine-tuning parameters 
indicates that the overall formulation is rather robust, independently of the details of the underlying biochemical reactions and absolute values of molecular concentrations. 
In contrast, modeling such a system using biochemical reaction equations requires
detailed information of the system,
including the absolute concentration of molecules, stoichiometries for each reaction, and a large number of biochemical reaction constants.
In our formulation, diagrams such as Fig.\ \ref{NF-kB}A encode all the dynamic information and provide an intuitive view of the system. 
Thus, our formulation can be more rapidly implemented, and is both robust and intuitive as compared with conventional biochemical reaction networks.
Clearly the validity of the approximations used here (FIG.\ \ref{illust}) 
is expected to be system-dependent and
must be investigated further.
Nevertheless, the linear and analytic nature of the formulation enables the use of several theoretical tools from various disciplines.
In addition, it is also interesting to note that intercellular interactions can be 
simultaneously implemented by letting the transition rate matrix depend on the probabilities of the nodes responsible for the intercellular activities.
Such a unified description of both intra- and inter- cellular interaction might help us to understand cell design in multicellular organisms,
though it is beyond the scope of this letter. 
We believe that this coarse-grained formulation of cellular systems provides a strong framework with which to integrate vast knowledge from molecular biology.

\section*{Acknowledgments}
We appreciate C. Furusawa and M. Sasai for valuable suggestions and comments.

\appendix
\section*{Appendix: The TNF--\NFkB model}
Tumor necrosis factor (TNF) is a soluble protein (cytokine). TNF is known to be involved in various biological processes such as apoptosis, inflammation, and immune responses \cite{aggarwal2003signalling}. Its intracellular signaling mechanism has been extensively studied \cite{hayden2008shared}. Several studies have built models of this system on the basis of biochemical reaction equations (reviewed in \cite{cheong2008understanding}), including that of Werner et al \cite{werner2008encoding} which we used for comparison with our formulation in this letter. 

We have assumed the topology of the system as shown in FIG.\ 3A in the main text.
This topology is much simpler than the biochemical reaction network diagram in \cite{werner2008encoding},
but is generally consistent with the current understanding of the TNF--\NFkB pathway.
The parameters needed for our modeling are the typical time-scales for each regulation.
We have adopted the appropriate parameters for rate-limiting steps in the data provided in \cite{werner2008encoding} and references therein
as typical time-scale parameters for our model.\\

\begin{table}[htbp]
\begin{tabular}{|rcc|c|c|c|}
\hline
\multicolumn{3}{|c|}{equation} & description&value&ref\\\hline
&$\stackrel{65}{\longrightarrow}$& !TNF  & The degradation rate of TNF in medium &0.0154 $\min^{-1}$&\cite{werner2008encoding}\\\hline
TNF &$\stackrel{3}{\longrightarrow}$& TNFR  & The net association rate, $k_{obs}$, of TNF and TNF-R1&$0.34 \min^{-1}$ &\cite{grell1998type}\\\hline
&$\stackrel{48}{\longrightarrow}$& !TNFR & The half-time $t_{\frac{1}{2}}$ of dissociation of TNF and TNF-R1&$33.2 \min$&\cite{grell1998type}\\\hline
TNFR &$\stackrel{5}{\longrightarrow}$& IKK & Assumed & &\\\hline
&$\stackrel{7}{\longrightarrow}$& !IKK & A fitted value for the inactivation rate of IKK&0.15 $\min^{-1}$&\cite{werner2008encoding}\\\hline
IKK &$\stackrel{3}{\longrightarrow}$& !\IkB & The IKK-mediated \IkB decay rate&0.36 $\min^{-1}$ &\cite{werner2008encoding}\\\hline
!\IkB &$\stackrel{0.2}{\longrightarrow}$& \NFkB &The \NFkB transportation rate to nuclei& 5.4 $\min^{-1}$&\cite{werner2008encoding} \\\hline
&$\stackrel{120}{\longrightarrow}$& \IkB & Assumed& & \\\hline
&&& Estimated by multiplying the &0.125 $\mu$M$\times$& \\
\IkB &$\stackrel{0.3}{\longrightarrow}$& !\NFkB &  characteristic concentration of \NFkB or \IkB&30 $\mu$M$^{-1}\min^{-1}$& \cite{werner2008encoding} \\
&&& by the association rate of \IkB and \NFkB && \\\hline
NFkB &$\stackrel{45}{\longrightarrow}$& \IkB & Assumed& & \\\hline
NFkB &$\stackrel{45}{\longrightarrow}$& A20 & Assumed& & \\\hline
A20 &$\stackrel{5}{\longrightarrow}$& !TNFR & Assumed& & \\\hline
&$\stackrel{345}{\longrightarrow}$& !A20 & The degradation rate of A20 protein&0.0029 $\min^{-1}$&\cite{werner2008encoding}\\
\hline
\end{tabular}\\
\caption{Equations and parameters of the TNF model}
\end{table}

The units of the above time-scale parameters are minutes.
There were five time-scale parameters which could not be obtained from the data provided in the references.
Though we could numerically fit them from the experimental data,
we chose to assume reasonable values for purpose of demonstration.
For example, the time-scales for inductions of \IkB and A20 by \NFkB were set to 45 $\min$
and the remaining time-scales were set to sum to roughly 10 $\min$
because the inductions of \IkB and A20 require approximately 1h.


\begin{thebibliography}{10}

\bibitem{de2002modeling}
H.~De~Jong.
\newblock Modeling and simulation of genetic regulatory systems: a literature
  review.
\newblock {\em Journal of computational biology}, 9(1):67--103, 2002.

\bibitem{edwards2001silico}
J.S. Edwards, R.U. Ibarra, and B.O. Palsson.
\newblock In silico predictions of escherichia coli metabolic capabilities are
  consistent with experimental data.
\newblock {\em Nature biotechnology}, 19(2):125--130, 2001.

\bibitem{hoffmann2002ikb}
A.~Hoffmann, A.~Levchenko, M.L. Scott, and D.~Baltimore.
\newblock The i$\kappa$b-nf-$\kappa$b signaling module: temporal control and
  selective gene activation.
\newblock {\em Science}, 298(5596):1241, 2002.

\bibitem{raj2008nature}
A.~Raj and A.~van Oudenaarden.
\newblock {Nature, nurture, or chance: stochastic gene expression and its
  consequences}.
\newblock {\em Cell}, 135(2):216--226, 2008.

\bibitem{tay2010single}
S.~Tay, J.J. Hughey, T.K. Lee, T.~Lipniacki, S.R. Quake, and M.W. Covert.
\newblock Single-cell nf-[kgr] b dynamics reveal digital activation and
  analogue information processing.
\newblock {\em Nature}, 2010.

\bibitem{kauffman1969metabolic}
S.A. Kauffman.
\newblock Metabolic stability and epigenesis in randomly constructed genetic
  nets.
\newblock {\em Journal of theoretical biology}, 22(3):437--467, 1969.

\bibitem{morris2010logic}
M.K. Morris, J.~Saez-Rodriguez, P.K. Sorger, and D.A. Lauffenburger.
\newblock Logic-based models for the analysis of cell signaling networks.
\newblock {\em Biochemistry}, 49(15):3216--3224, 2010.

\bibitem{kaern2005stochasticity}
M.~K{\ae}rn, T.C. Elston, W.J. Blake, and J.J. Collins.
\newblock {Stochasticity in gene expression: from theories to phenotypes}.
\newblock {\em Nature Reviews Genetics}, 6(6):451--464, 2005.

\bibitem{bala'zsi2011cellular}
G.~Bal{\'a}zsi, A.~van Oudenaarden, and J.J. Collins.
\newblock Cellular decision making and biological noise: From microbes to
  mammals.
\newblock {\em Cell}, 144(6):910--925, 2011.

\bibitem{shmulevich2002probabilistic}
I.~Shmulevich, E.R. Dougherty, S.~Kim, and W.~Zhang.
\newblock Probabilistic boolean networks: a rule-based uncertainty model for
  gene regulatory networks.
\newblock {\em Bioinformatics}, 18(2):261, 2002.

\bibitem{gillespie1977exact}
D.T. Gillespie.
\newblock {Exact stochastic simulation of coupled chemical reactions}.
\newblock {\em The journal of physical chemistry}, 81(25):2340--2361, 1977.

\bibitem{gardner2000construction}
T.S. Gardner, C.R. Cantor, and J.J. Collins.
\newblock Construction of a genetic toggle switch inescherichia coli.
\newblock {\em Nature}, 403:339--342, 2000.

\bibitem{ferrell2011modeling}
J.E. Ferrell~Jr, T.Y.C. Tsai, and Q.~Yang.
\newblock Modeling the cell cycle: Why do certain circuits oscillate?
\newblock {\em Cell}, 144(6):874--885, 2011.

\bibitem{werner2008encoding}
S.L. Werner, J.D. Kearns, V.~Zadorozhnaya, C.~Lynch, E.~OfDea, M.P. Boldin,
  A.~Ma, D.~Baltimore, and A.~Hoffmann.
\newblock Encoding nf-$\kappa$b temporal control in response to tnf: distinct
  roles for the negative regulators i$\kappa$b$\alpha$ and a20.
\newblock {\em Genes \& development}, 22(15):2093, 2008.

\bibitem{aggarwal2003signalling}
B.B. Aggarwal.
\newblock Signalling pathways of the tnf superfamily: a double-edged sword.
\newblock {\em Nature Reviews Immunology}, 3(9):745--756, 2003.

\bibitem{hayden2008shared}
M.S. Hayden and S.~Ghosh.
\newblock Shared principles in nf-[kappa] b signaling.
\newblock {\em Cell}, 132(3):344--362, 2008.

\bibitem{cheong2008understanding}
R.~Cheong, A.~Hoffmann, and A.~Levchenko.
\newblock Understanding nf-$\kappa$b signaling via mathematical modeling.
\newblock {\em Molecular systems biology}, 4(1), 2008.

\bibitem{grell1998type}
M.~Grell, H.~Wajant, G.~Zimmermann, and P.~Scheurich.
\newblock The type 1 receptor (cd120a) is the high-affinity receptor for
  soluble tumor necrosis factor.
\newblock {\em Proceedings of the National Academy of Sciences}, 95(2):570,
  1998.

\end{thebibliography}

\end{document}